%% file: main.tex
\documentclass[conference]{IEEEtran}
\IEEEoverridecommandlockouts

\usepackage[linesnumbered,ruled,vlined]{algorithm2e}
\usepackage{pdfpages}
\AtBeginDocument{%
  }

\usepackage{amsfonts}
\usepackage{booktabs}
\usepackage{multirow}
\usepackage{graphicx}
\usepackage{booktabs} 
\usepackage{makecell}
\usepackage{comment}  
\usepackage{url}
\usepackage{amsmath}  
\usepackage{amssymb}  
\usepackage{xcolor}
\usepackage{subcaption}
\usepackage{subcaption}
\definecolor{atkred}{RGB}{192, 0, 0}
\definecolor{defblue}{RGB}{0, 112, 192}
\definecolor{darkgreen}{RGB}{18, 124, 19} 
\ifCLASSOPTIONcompsoc
  \usepackage[nocompress]{cite}
\else
  \usepackage{cite}
\fi

\begin{document}
%
\title{Developing a Strong CPS Defender: An Evolutionary Approach}


\author{
    \IEEEauthorblockN{
        Qingyuan Hu\IEEEauthorrefmark{2},
        Christopher M. Poskitt\IEEEauthorrefmark{3},
        Jun Sun\IEEEauthorrefmark{3},
        Yuqi Chen$^{*}$\thanks{* Corresponding author.}\IEEEauthorrefmark{2}
    }
    \IEEEauthorblockA{
    \IEEEauthorrefmark{2}ShanghaiTech University, China, \{huqy2022, chenyq\}@shanghaitech.edu.cn;
    \\
    \IEEEauthorrefmark{3}Singapore Management University, Singapore, \{cposkitt, junsun\}@smu.edu.sg}
}

\maketitle

\begin{abstract} 
    Cyber-physical systems (CPSs) are used extensively in critical infrastructure, underscoring the need for anomaly detection systems that are able to catch even the most motivated attackers. Traditional anomaly detection techniques typically do `one-off' training on datasets crafted by experts or generated by fuzzers, potentially limiting their ability to generalize to unseen and more subtle attack strategies. Stopping at this point misses a key opportunity: a defender can actively challenge the attacker to find more nuanced attacks, which in turn can lead to more effective detection capabilities. Building on this concept, we propose \emph{Evo-Defender}, an evolutionary framework that iteratively strengthens CPS defenses through a dynamic attacker-defender interaction. \emph{Evo-Defender} includes a smart attacker that employs guided fuzzing to explore diverse, non-redundant attack strategies, while the self-evolving defender uses incremental learning to adapt to new attack patterns. We implement \emph{Evo-Defender} on two realistic CPS testbeds: the Tennessee Eastman process and a Robotic Arm Assembly Workstation, injecting over 600 attack scenarios. 
    In end-to-end attack detection experiments, \emph{Evo-Defender} achieves up to 2.7× higher performance than state-of-the-art baselines on unseen scenarios, while utilizing training data more efficiently for faster and more robust detection.
\end{abstract}


\IEEEpeerreviewmaketitle

\input{content/introduction}

\input{content/background}

\input{content/approach_and_implementation}

\input{content/evaluation}

\input{content/related_work}

\input{content/Conclusions}

\vspace{-0.2em}
\section*{Acknowledgment}
\vspace{-0.2em}
We sincerely appreciate the anonymous reviewers for their valuable feedback, which significantly enhanced this paper. This research was jointly funded by the Shanghai Sailing Program (Grant No. 23YF1427500), the NSFC Program (Grant No. 62302304), and the ShanghaiTech Startup Funding.



\bibliographystyle{IEEEtran}
\bibliography{sample-base}

\end{document}

%% file: content/introduction.tex
\section{Introduction}\label{sec:intro}

Cyber-physical systems (CPSs), which tightly integrate cyber and physical components, are widely deployed in critical domains such as water treatment, electricity distribution, and pharmaceutical manufacturing.
High-profile incidents---including the Stuxnet-induced centrifuge failure at Natanz~\cite{stuxnet} and the Industroyer-triggered grid collapse in Kyiv~\cite{blackhat2017}---underscore how CPS vulnerabilities can translate into physical-world catastrophes.
The inherent complexity and legacy nature of many CPSs creates security blind spots that traditional IT defenses struggle to address, making it challenging to build robust protection mechanisms.
Consequently, anomaly detection has become a widely relied-upon strategy, particularly for legacy systems. 
While mathematically modeling CPS behavior for anomaly detection is often challenging and time-consuming, the abundance of sensor and actuator data has led to the widespread adoption of machine learning algorithms to develop data-driven, time-series anomaly detectors \cite{anomaly-detect-unsupervised-learning-yuqi, Truth_Will_Out, Power_Grid_Controller_Anomaly_Detection, Detecting_Cyber_Attacks_in_ICS, TABOR, Learning_Based_Anomaly_Detection_for_Industrial_Arm_Applications, High_Performance_Unsupervised_Anomaly_Detection_for_Cyber_Physical_System_Networks}. 

For data-driven approaches, detection performance heavily depends on the quality and diversity of the training dataset.
Unfortunately, these datasets typically only capture benign operating conditions, limiting their utility for detecting real-world attacks.
Collecting data from scenarios involving active attacks or abnormal behavior---such as those in \cite{itrust_dataset_2025, swatdata}---requires significant time, effort, and domain expertise.
Even then, such datasets can carry inherent biases.
Furthermore, generalizing these methods across diverse CPSs---particularly legacy systems---remains highly impractical due to each system's unique processes, mechanisms, and complexities.
This creates a fundamental paradox: defenders not only need vast amounts of high-quality data to achieve robust detection, but also require substantial computational and storage resources to process and retrain on such data. However, in practice, they often lack both, making effective and adaptive defense especially challenging.

To address this challenge, prior work has applied automated testing and fuzzing techniques to systematically explore the behavior of CPSs~\cite{learning-guided-network-fuzzing-for-testing-cps, active-fuzzing}.
These methods can efficiently generate test cases that drive the system into hazardous states, making them useful for evaluating existing anomaly detection mechanisms.
However, they are less effective for producing the clean, diverse data required to train robust anomaly detectors.
These techniques typically rely on a machine learning model trained on sensor and actuator logs or network traffic to guide the automated discovery of diverse attacks targeting different sensors or actuators.
Their key limitation lies in their singular focus on achieving attack goals, without consideration for co-developing defensive strategies.
Consequently, while these approaches may partially address data scarcity, they often neglect data quality and generate redundant or noisy data. This undermines their usefulness for building robust defenders, which require significant computational resources.

\begin{figure*}[t]
    \centering
    \includegraphics[width=1\linewidth]{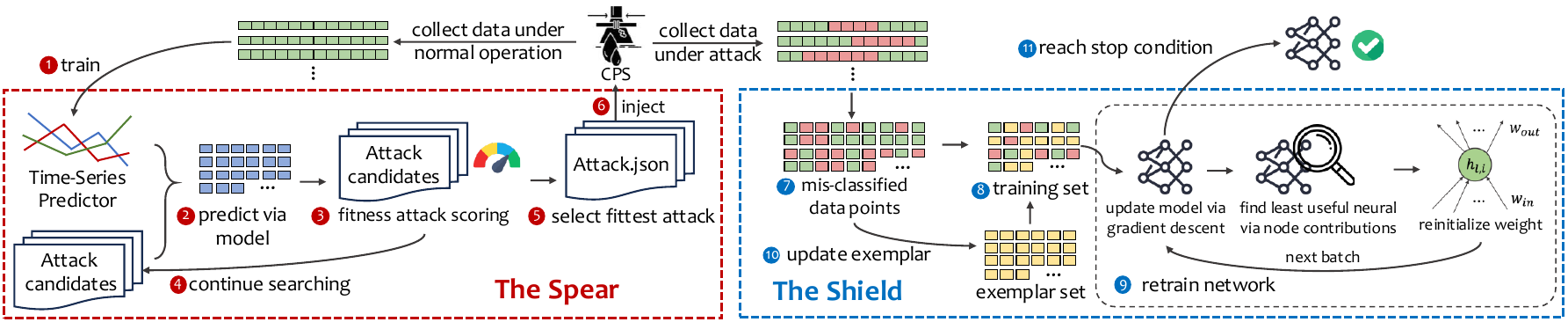}
    \vspace{-1.5em}
    \caption{Overview of our approach}
    \vspace{-1.8em}
    \label{fig:main_arch}
\end{figure*}

To address these challenges, we propose a novel framework, \textit{Evo-Defender}, designed to automatically construct a robust CPS attack detection system, particularly for scenarios where the defender has access to limited data due to constrained computing power and memory. Our framework consists of two main components: the Spear and the Shield, as illustrated in Figure \ref{fig:main_arch}.
In the first component, the objective is to extract data from the CPS through a diverse range of attacks, capturing various abnormal states the system may encounter. To mitigate the challenges posed by the massive search space and data homogenization, we employ a machine learning model to guide the Spear, ensuring it avoids redundant attack strategies while achieving multiple attack objectives. By exposing the CPS to unsafe states, this process reveals critical attack behaviors, thereby strengthening the Shield.
In the second component, we implement an incremental learning approach tailored to the time-series data generated by the Spear. This method addresses both the scarcity of abnormal CPS state data and the resource constraints by supporting dynamic model updates without retraining from scratch or storing vast historical logs. 
This enables efficient knowledge accumulation, rapid adaptation, and reduces issues with redundant or low-quality data scenarios typical of CPS environments.
First, we filter out misclassified data from the latest CPS logs using the current defender, which is typically a machine learning-based anomaly detector initialized randomly. Then, a multi-module incremental learning strategy—incorporating exemplars, imbalanced learning, and continual backpropagation—leverages the misclassified data to enhance the Shield's performance.
This approach minimizes the need for extensive data storage and retraining while rapidly integrating newly discovered threat patterns into the defense, eliminating the need to wait for a complete attack dataset before improving detection capabilities.

We evaluated the effectiveness of our solution using two realistic CPS testbeds: the Tennessee Eastman process (TE) and the Robotic Arm Assembly Workstation (RAAW).
TE is a widely used chemical process simulation environment developed by Matlab Simulink\cite{TE-simulink}, whereas RAAW is a real-world robotic arm platform.
We deployed \textit{Evo-Defender} on these platforms, injected 336 and 300 attacks respectively, and recorded the entire process in detail for the training and evaluation.
For TE, the injected attacks targeted high reactor pressure, high temperatures, and abnormal tank liquid levels, whereas for RAAW, attacks attempted to halt CPS operations and cause a sudden pause in movement.
With incremental training, \textit{Evo-Defender} achieves 92.03\% and 89.59\% accuracy on two platform test sets, improving upon the best baselines by 23.38\% and 29.07\%, respectively.
In end-to-end tests, the model correctly identifies 89 of 103 and 38 of 44 attacks on their respective test sets. Compared to the best baseline, the TE platform’s false positive rate drops to one-third, while the RAAW platform’s detection rate is nearly tripled.
Finally, we conducted an ablation study on \textit{Evo-Defender} to assess each module's contribution across various incremental learning module configurations. 
The results indicated that our method requires fewer evolution rounds than other learning schemes and utilizes data more efficiently, leading to faster improvements in defender's performance.

In summary, our work contributes the following:
\begin{itemize}
\item \textbf{Automated Attack Generation}: We develop a smart attacker that efficiently uncovers varied attack trajectories, generating a comprehensive dataset of over 600 attack scenarios on two real CPS platforms.

\item \textbf{Continual Learning Defender}: We propose the first online scheme for incrementally updating the defense model, with the goal of improving existing deep learning-based anomaly detection methods when faced with incorrect incoming prediction data. In end-to-end testing scenario, our defender outperformed the baseline by at least 2.7 times in accuracy. In terms of data consumption, our approach used 69.6\% less data and delivered higher accuracy than the baseline.

\item \textbf{Robust Detection Capability}: We implement our approach on two CPS platforms, demonstrating its ability to detect a wide range of unsafe states and validating the effectiveness of continual evolution in addressing newly emerging threats. 

\item \textbf{Data Availability}: We provide the results at \cite{evo-defender-artifacts}.
\end{itemize}

The remainder of this paper is organized as follows: Section \ref{sec:background} introduces the background of our two CPS testbeds and problem definition. 
Section \ref{sec:approach} details the architecture of our \textit{Evo-Defender} framework and a case study demonstrating the deployment of our approach on a realistic CPS platform. 
Section \ref{sec:eval} provides experimental evaluation across accuracy and robustness metrics. Section \ref{sec:related} reviews related work in adversarial machine learning for CPS. 
The paper concludes with final remarks in Section \ref{sec:conclusion}.

%% file: content/background.tex
\begin{figure}[t]
    \centering
    \includegraphics[width=0.95\linewidth]{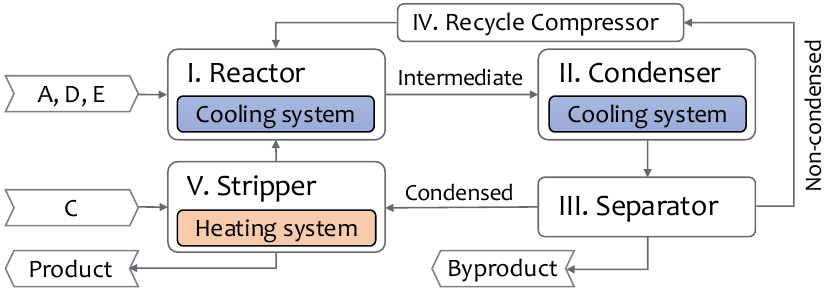}
    \vspace{-0.3em}
    \caption{The Tennessee Eastman Challenge Process. A, C, D, and E are the are the four raw materials participating in the reaction}
    \vspace{-1em}
    \label{te_arch}
\end{figure}

\begin{figure}[t]
    \centering
    \includegraphics[width=1\linewidth]{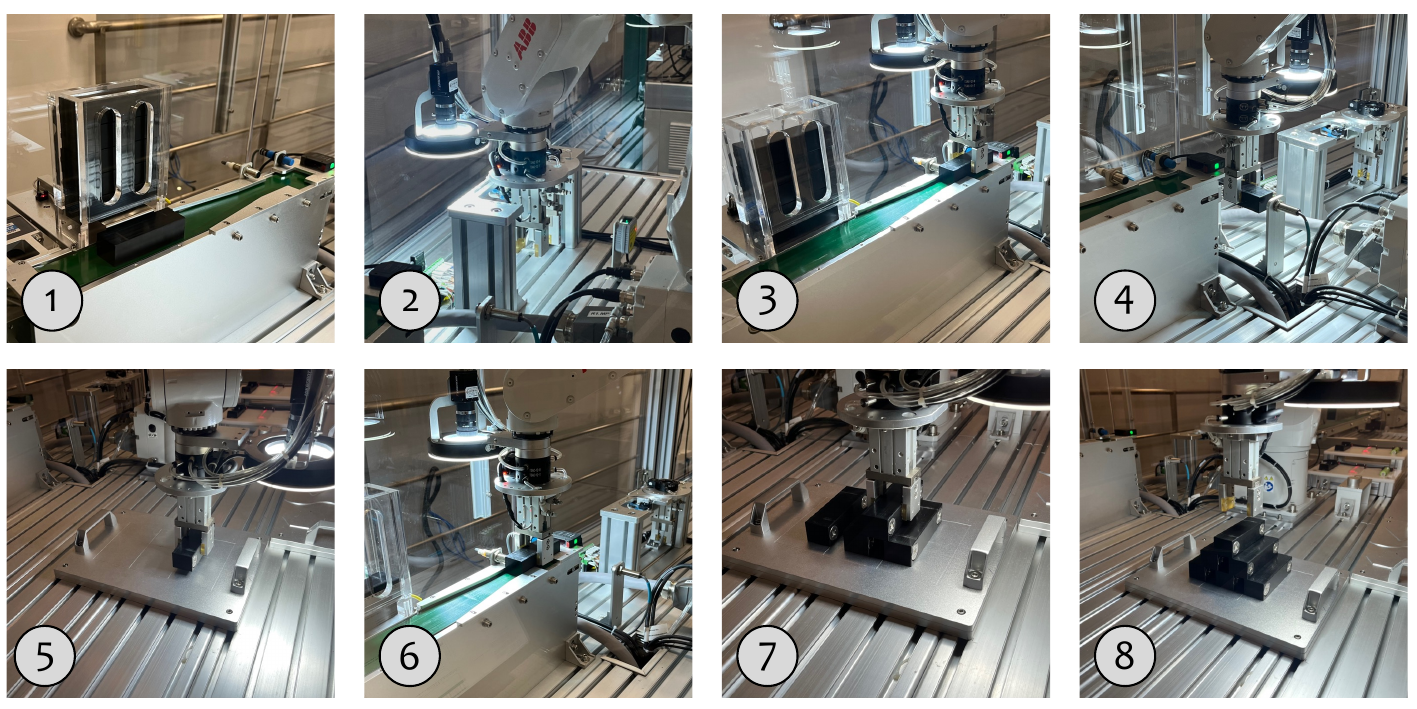}
    \vspace{-1.5em}
    \caption{The workflow of a robotic arm in palletizing tasks}
    \vspace{-1.8em}
    \label{robot_workflow}
\end{figure}

\begin{figure}[t]
    \centering
    \includegraphics[width=0.85\linewidth]{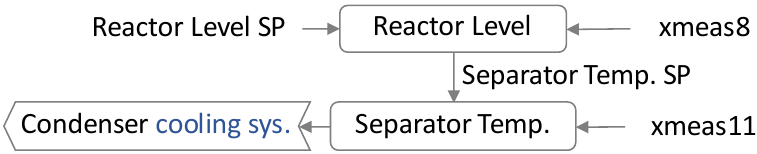}
    \vspace{-0.4em}
    \caption{Example of a cascading controller in TE}
    \vspace{-1.5em}
    \label{fig:cascading_control}
\end{figure}

\vspace{-0.2em}
\section{Background and Problem Definition}\label{sec:background}
\vspace{-0.2em}
\label{sec:background}

In this section, we provide an overview of the two CPS platforms—TE and RAAW—which serve as the testbeds for evaluating our approach. We then present a problem definition followed by a summary of the threat model considered in this work.

\subsubsection*{\emph{\textbf{TE Testbed}}}
The Tennessee Eastman (TE) process is a computational model that simulates a chemical plant with five main units: a reactor, condenser, compressor, separator, and stripper. 
It is widely used for designing and testing control algorithms, with a well-documented architecture and implementation\cite{TE-simulink, Plant-wide-control-of-the-Tennessee-Eastman-problem}.
As shown in Figure \ref{te_arch}, the process involves reactants (A, C, D, E) converted into two products (G, H), along with an inert and a byproduct, after passing through the main units. 
Fung's work \cite{attribution-icsanomaly-ndss2024} provides modified TE code for sensor and actuator manipulations implemented in Matlab Simulink, making it an ideal testbed for our modification and testing.
We built upon their work to enable more flexible manipulations, including 25 sensors, 9 actuators, and 17 PID controller configurations.

The TE design includes defining safe operational states through specific constraints that are vital for equipment protection and safety.
Any violation of these constraints triggers an automatic system shutdown. 
Downs and Fogel~\cite{Plant-wide-control-of-the-Tennessee-Eastman-problem} outline specific operational constraints for each process unit. 
For example, if the Reactor Pressure indicator detects a pressure exceeding 3000 kPa, the system will execute its interlock strategy to shut down the process.

\subsubsection*{\emph{\textbf{RAAW Testbed}}}
RAAW is a fully automated platform featuring an ABB robotic arm and a Siemens S7-1200 PLC, creating a two-node distributed CPS capable of executing a wide range of assembly tasks.
The PLC uses 29 digital inputs and 22 digital outputs to communicate with sensors, actuators, and the robotic arm, facilitating the exchange of states and signals.
It interacts with components like warehouse sensors, conveyor belt sensors, gripper position sensors, and safety door sensors.
In this study, we assign it a palletizing task: stacking six plastic rectangular prisms into a pyramid-shaped structure.

As shown in Figure \ref{robot_workflow}.1, the process begins with a plastic material being pushed from the warehouse to specific locations via a conveyor belt. 
Simultaneously, the robotic arm retrieves its gripper (\ref{robot_workflow}.2).
The conveyor belt transports the pallet to the designated location, where the robotic arm grabs the material (\ref{robot_workflow}.3). 
The robotic arm uses sensors to ensure that it has successfully captured the material (\ref{robot_workflow}.4). 
After grasping the material, the robotic arm places it on the platform (\ref{robot_workflow}.5). 
This action is repeated as the robotic arm retrieves additional materials (\ref{robot_workflow}.6-7). 
Finally, it positions the last material to complete the pyramid formation (\ref{robot_workflow}.8).
Through six rounds of PLC-triggered grasping and placement, six plastic materials are built into a pyramid.

The PLC stores 127 variables in fixed memory locations, similar to programming variables, to facilitate palletizing and management of the alarm system.
Similar to TE, RAAW has predefined safety limits.
For example, if a safety door is open, an alarm triggers, setting a Boolean variable to true, which activates the alarm procedure.

\subsubsection*{\emph{\textbf{Problem Definition}}}
A CPS is a decentralized reactive system, interacting with its physical environment via inputs and outputs in an ongoing manner.
Consider the TE testbed as an example.  Figure~\ref{fig:cascading_control} illustrates a cascading control structure designed to maintain the reactor's liquid level. Rather than controlling the reactor level directly, the system achieves regulation by manipulating downstream components. Specifically, a master control loop continuously monitors the liquid level via sensor \textit{xmeas8} and generates a Separator Temperature Setpoint by combining the measured value with a predefined Reactor Level Setpoint. A slave control loop then controls the separator temperature by processing the setpoint from the master loop and the temperature signal from sensor \textit{xmeas11}, thereby managing the condenser’s cooling system. The TE system comprises numerous such interdependent controllers, all working toward a common operational objective. This complexity introduces significant system analysis challenges, especially when attempting to reason about behavior under adversarial conditions.

Given the tightly coupled and domain-specific nature of CPSs, both attacking and defending them demand not only cybersecurity expertise but also a deep understanding of the underlying physical processes.
To reduce reliance on such domain knowledge and to lower the associated costs across diverse CPS platforms, we draw inspiration from prior work on fuzz testing~\cite{fuzzing-for-software-security-testing-and-quality-assurance} to uncover attacks that induce abnormal system states by generating malicious network packets or control commands.
However, due to the inherent nature of fuzzing, the generated data often exhibit low quality, class imbalance, and high redundancy. Moreover, real-world CPS deployments face challenges such as sensor drift~\cite{sensor-drift-1, sensor-drift-2} in physical processes, which cause ongoing shifts in data distribution. These factors result in dynamically evolving data characteristics, which limit the effectiveness of static, batch-trained anomaly detectors. Therefore, incremental learning is essential in our problem setting, as it enables continual model adaptation to both the heterogeneous fuzzing-generated samples and real-world changes, ultimately improving the robustness and coverage of the anomaly detector.

\subsubsection*{\emph{\textbf{Threat Model}}}
As the objective of the Spear is to generate data that can be used to enhance and evaluate \textit{Evo-Defender}, we first clarify the assumptions made regarding the CPS and the nature of potential attacks.
In our approach, the Spear is designed to emulate real-world adversaries who target control components and the communication channels between sensors and actuators, thereby enabling network-based attacks.

We assume that adversaries possess certain privileges within the CPS, enabling them to access and modify actuator commands, sensor readings, and system configurations. Additionally, attackers are assumed to have access to ground-truth sensor data, allowing them to observe the effects of their manipulations. Our threat model accommodates a spectrum of adversaries, ranging from highly capable attackers—who can simultaneously compromise multiple components—to more constrained ones with limited access (e.g., only a subset of sensors). While this model may appear to empower the attacker, our objective is not to simulate a likely real-world compromise, but rather to construct a diverse and systematic framework for evaluating the resilience of defensive mechanisms. By enabling the simulation of a wide range of adversarial behaviors, this flexible framework facilitates comprehensive testing and contributes to the development of more robust defensive strategies within \textit{Evo-Defender}.

%% file: content/approach_and_implementation.tex
\vspace{-0.2em}
\section{Approach and Implementation}\label{sec:approach}
\vspace{-0.2em}

The overall objective of our solution is to quickly develop a robust defender without depending on existing datasets or requiring extensive computing resources.
To achieve this goal, we propose a practical, generalizable framework that leverages the Spear to continuously reinforce the Shield.
Our approach enables the creation of a powerful defender with minimal data while maintaining high accuracy in identifying real-world attacks.

Our approach mainly consists of two components, the Spear and the Shield, as illustrated in Figure \ref{fig:main_arch}.
In the Spear, we train a predictive model using data from a CPS historian, which includes sensor readings, actuator values, and configurations.
This guides the Spear in identifying and injecting diverse effective attacks into the CPS to observe its operational state under attack.
Subsequently, we apply incremental learning techniques using these operational states to enhance the defender's capabilities, allowing it to learn the latest attack characteristics and update its defense strategies.
Finally, through iterative execution of these steps, we develop a robust defender and validate its performance with real testing datasets.

In the following, we outline the general implementation of these broad steps and illustrate their application using the TE testbed as an example, with the same approach also applicable to the RAAW platform. The corresponding code is provided in the supplementary materials \cite{evo-defender-artifacts}.

\vspace{-0.4em}
\subsection{Pre-Requisites}
\vspace{-0.2em}

The Spear stage requires a model capable of predicting the effects of manipulating sensor and actuator inputs. This is a standard prerequisite for CPS-based testing (e.g., \cite{learning-guided-network-fuzzing-for-testing-cps,active-fuzzing}), and we elaborate on the steps below.

\subsubsection*{\emph{\textbf{Data Collection}}}
Training an effective prediction model requires a dataset from the target CPS that captures the relationships between actuator values, sensor readings, controller configurations, and future system states. This data should be recorded at regular intervals and encompass a range of operational scenarios to ensure comprehensive coverage.
The dataset can be obtained from the system historian or collected passively by monitoring SCADA \cite{SCADA} communications or network traffic, provided the CPS operates without external interference.
To further enrich the data and expose additional system behaviors, we may also introduce controlled variations in configurations or actuator settings during runtime.

\emph{Implementation for TE.}
We utilized the TE process in MATLAB Simulink to achieve this goal, incorporating both data recording and real-time configuration adjustments.
Each operational cycle was simulated over 12 hours, during which we recorded data points every 0.5 seconds.
To increase data variability, we introduced several manually crafted attacks after 5 hours of runtime, such as altering a PID controller configuration to a random value, and continued system operation until a timeout or shutdown occurred. 
After each run, we reset the system for the next data collection round. 
In total, we recorded 30 data entries, each containing between 10,000 and 24,000 sampling points, including sensor readings, actuator values, and configuration parameters. The dataset was randomly partitioned, with 80\% allocated for training and the remaining 20\% reserved for model validation.\\
\vspace{-1.2em}
\subsubsection*{\emph{\textbf{Training a Prediction Model}}}
In this step, we use the collected raw data to train a model that guides the Spear to identify effective manipulations.
Before training, we convert the well-organized dataset into a series of vectors with a fixed format suitable for processing by a learning algorithm, such as a neural network.
In this work, we assume that the status of the CPS is described by time series data at discrete time points, represented by three kinds of variables: sensor readings, actuator values, and configurations. 
These values can be either discrete or continuous.
Based on this assumption, a possible form is $\langle \textbf{A}_{0:n}, \textbf{S}_{0:n}, \textbf{C}_{0:n} \rangle$, where the $\textbf{A}_t$ $\textbf{S}_t$ and $\textbf{C}_t$ signify the sets of values for actuators, sensors, and system variables, respectively, at time point $t$.
Given that this vector includes both continuous and discrete values, an appropriate data normalization method must be applied after data collection. Once the feature vectors are defined, we proceed to train a supervised machine learning model to predict future features. It is important to note that while the model should ideally be accurate with respect to observed data, there is flexibility in the required level of accuracy. Since the attacker seeks to identify new attacks for which no prior data exists, the model does not need to be perfectly accurate for all unseen scenarios. Instead, the model is expected to capture the correct trends, enabling it to guide the identification of effective manipulations. Furthermore, models can be refined over time to address these blind spots, as demonstrated in related works such as \cite{active-fuzzing}.

\emph{Implementation for TE.}
We assume that the Spear manipulates CPS through the network, and the future state of TE depends on its previous operational state.
Therefore, the input format is $\langle f_0, f_1, ..., f_t \rangle$, where $f_t$ represents the feature at time point $t$, $f_t = \langle s_{0,t}, ..., s_{n,t}, a_{0,t},..., a_{m,t}, c_{0,t},...,c_{l,t} \rangle$ for each time point $t$.
As our evaluation of attack success is based on violations of the TE system’s operational constraints, we focus our prediction on a subset of critical sensor readings directly tied to these constraints—specifically, reactor pressure, reactor level, reactor temperature, separator level, and stripper level.

For model selection, we employ a Long Short-Term Memory (LSTM) network due to its effectiveness in handling sequential and multivariate data, as well as its capability to capture complex nonlinear dependencies—characteristics essential for modeling CPS dynamics. Our implementation utilizes a straightforward LSTM architecture followed by a linear output layer. 

\vspace{-0.4em}
\subsection{The Spear}
\vspace{-0.2em}
\input{algorithm/algorithm-gen-attack-vector} 
In the Spear, our objective is to systematically find attack vectors that the existing anomaly detection model cannot catch. To achieve it, building on Chen et al.'s work \cite{learning-guided-network-fuzzing-for-testing-cps, active-fuzzing}, we use random search and genetic algorithms to generate meaningful attack vectors, as illustrated in Algorithm~\ref{algo:gen-attack-vector}. As outlined in the threat model, we assume that the Spear can arbitrarily manipulate components within the CPS through network-based attacks (e.g., man-in-the-middle attacks). We represent a sequence of such attacks using a floating-point vector, where each element corresponds to a modification applied to a specific variable within the CPS under test.

Unlike previous work that focuses solely on finding more attacks, we place greater emphasis on attack diversity to maximize the coverage of the attack surface. Existing CPS testing approaches typically pursue a single objective, limiting exploration and making it difficult to uncover more system vulnerabilities. Our design encourages the Spear to generate diverse attacks, using a fitness function to evaluate the similarity between new and historical attacks. For example, if most existing attacks have caused an increase in the reactor tank’s liquid level, subsequent attacks should aim to trigger other system behaviors in the CPS, with higher scores assigned to more distinct effects. In addition, we focus on generating attacks that significantly deviate the system from its normal operational state; the fitness function also considers actual sensor readings, assigns higher scores to attacks causing greater deviations  from the normal sensor value ranges.

By integrating these two design considerations---behavioral diversity and deviation from normalcy---the fitness function effectively steers the search process toward the generation of diverse and impactful attack vectors. Once formally defined, the fitness function assigns a fitness value to each attack vector, reflecting its threat level (Algo \ref{algo:gen-attack-vector}, Line 5). To refine these attack vectors, we can apply heuristic search strategies such as random search (Algo \ref{algo:gen-attack-vector}, Lines 2–5) or genetic algorithms (Algo \ref{algo:gen-attack-vector}, Lines 6–9) to explore the space of sensor values, actuator states, and system configurations. 
After generating a set of attack candidates, we use Roulette Wheel Selection \cite{roulette_wheel_selection} to choose the attack. Each candidate's selection probability is proportional to its fitness, calculated as \(f_i / \sum_{j=1}^n f_j\), where \(f_i\) is the fitness of candidate \(i\). We generate a random number between 0 and the total fitness sum, then iterate through the candidates, accumulating their fitness values until the sum exceeds the random number. The candidate at this point is selected as the attack vector.
This approach ensures both the diversity and effectiveness of the generated attack vectors in evaluating the security and robustness of the CPS.

Once the attack vectors are generated, we inject them into the CPS to evaluate the performance of the defense mechanism, specifically targeting manipulations that can drive certain system properties into unsafe states without triggering alarms from the anomaly detector. Whether or not data is logged depends on the response of the defender. If the defender fails to detect the injected attack, or erroneously raises an alarm when the CPS is operating normally (i.e., a false positive), we record the corresponding data along with its generated label for that period. If the system continues to operate normally until a predefined time limit is reached, we consider the attack unsuccessful, and the resulting data is labeled as normal. 
After each injection, the system is reset to normal operation, and this process repeats until the defender meets a predefined stopping condition (e.g., when the defender consistently detects attacks exceeding a specified threshold multiple times).

\input{algorithm/coverage-guided}
\emph{Implementation for TE.}
In TE, all sensors, actuators, and configuration parameters are represented as floating-point values, allowing attacks---defined as sequences of control commands---to be encoded as float vectors.
Each element in an attack vector corresponds to a command that alters a specific configuration.
For example, an attack vector $\langle -2.0, 0.5, \ldots, 0.0 \rangle$ represents a series of commands aimed at causing a reactor overflow. 
The first element changes the reactor temperature controller's proportional gain (Kc) from -8 to 8, the second updates the integral gain (Ti) parameter from 0.125 to 0.1875, and so on, with 0.0 indicating no change to a configuration variable.
In this case, the attack changed the proportional gain from -8 to +8, causing the controller to issue heating commands rather than cooling commands as the temperature rose. This led to a rapid temperature increase, intense reactor vaporization, and a pressure surge that triggered the high-pressure alarm.
For manipulating sensors and actuators, we follow the method described by Fung et al.~\cite{attribution-icsanomaly-ndss2024} to apply diverse attack approaches.

The fitness function in TE consists of 2 parts.
The first is a coverage-guided fitness function (see Algorithm~\ref{algo:coverage-guided}), which encourages the exploration of diverse vulnerabilities rather than focusing solely on the easiest ones to find.
In our LSTM-based predictor, the process typically begins with the output of an embedding, which is then passed through a fully connected layer to generate the final prediction.
We save the embeddings from the historian attack vectors to create an embedding set (Algo~\ref{algo:coverage-guided}, Lines 2-4), which is then used to evaluate each candidate during each round of search process.

In the second part of the fitness function, we define it as follows:
\[
    f(v_s) = 
    \begin{cases}
        1 - \frac{\min \left((v_{s}-L_{s}),(H_{s}-v_{s})\right)}{H_s - L_s}, & \text{if } v_{s} \in [L_{s}, H_{s}] \\
        1 + \frac{\min \left(\left|v_{s}-L_{s}\right|,\left|v_{s}-H_{s}\right|\right)}{H_s - L_s},  & \text{otherwise}
    \end{cases}
\]
where $v_s$ is the value of the current sensor s, $L_s$ is its lower threshold, and $H_s$ is the upper threshold, and $H_s - L_s$ represents the safety range of the sensor.
The final fitness value for an attack, calculated by summing the two fitness components described above, measures how close sensor values are to safety limits, guiding the search for both effective and unexplored attack vectors.

In our study, we configure the random search algorithm with a population size of 100. 
For the genetic algorithm, we set the population size to 100, and used crossover and mutation to generate 20 offspring.
These parameters enable the Spear to efficiently determine an attack vector within 10 seconds.

\vspace{-0.4em}
\subsection{The Shield}
\vspace{-0.2em}

The Shield can be divided into two stages: first, collecting misclassified data, and then using it for incremental learning to strengthen the defender.

\subsubsection*{\emph{\textbf{Collect misclassified data}}}
In this component, as illustrated in Figure~\ref{fig:main_arch}, incoming data traces---potentially affected by attacks---are evaluated by the current defender, which may initially be a randomly initialized machine learning classifier. If the defender produces an incorrect judgment, the misclassified data points within the trace are identified and filtered. These points are then incorporated into the exemplar set to retrain the defender using incremental learning techniques. 

In addition, it is worth noting that sensor drift~\cite{sensor-drift-1, sensor-drift-2}—a gradual, linear increase in sensor output—is a common issue in real-valued sensor nodes. If this drift exceeds the tolerance threshold of the CPS system and results in abnormal behavior, the Shield should be able to detect it. Such conditions can be simulated using the attack vector generation strategy described in the previous section. Conversely, if the drift remains within the system's tolerance range, the Shield should refrain from triggering alarms to prevent unnecessary disruption to CPS operations. To simulate sensor drift, a slight offset can be encoded into an attack vector and injected during CPS initialization.

To gather data points for reinforcing the Shield, we first employ the current model to assess incoming traces. This evaluation uses an anomaly detector applied to the entire trace via a sliding window approach. The Shield may exhibit the following types of erroneous judgments:
\begin{itemize}
    \item Prematurely raising an alarm before the attack is injected,
    \item Failing to raise an alarm after a valid attack is injected but before system shutdown (false negative),
    \item Incorrectly raising an alarm following an ineffective attack before the predefined time limit (false positive).
\end{itemize}

These misclassified instances are extracted and subsequently incorporated into the retraining process to incrementally enhance the Shield’s performance over time.

\emph{Implementation for TE.}
When the Spear generates an attack that is misclassified by the anomaly detector, we record the corresponding data log and transform it into a feature vector formatted as $\langle f_0, f_1, \ldots, f_t \rangle$. Here, $f_t = \langle a_{0,t}, \ldots, a_{n,t}, s_{0,t}, \ldots, s_{m,t} \rangle$ denotes the actuator and sensor values at time $t$, while $l_t$ indicates whether the CPS state at that time is normal or abnormal.
To simulate sensor drift, we inject attacks by targeting up to 10 sensors or actuators per injection during the fuzzing process. Offsets are randomly assigned to these targets, with magnitudes constrained to no more than four times their standard deviation under normal operating conditions. This constraint ensures that the simulated drift remains within the TE system’s operational tolerance.

The current anomaly detector is then used to analyze the entire trace using a sliding window approach. If the detector produces incorrect judgments, the corresponding traces are saved for use in improving the model. In total, we injected and generated 223 attacks and associated data entries for enhancing the detector. An additional 113 attacks were used to evaluate the performance improvements. A detailed analysis of this dataset will be provided in the evaluation section.

\subsubsection*{\emph{\textbf{Incremental learning}}}
In this step, the Shield leverages the misclassified samples collected previously to further refine its model and enhance its defensive capabilities. The rationale is that the detection model should be promptly updated whenever misclassified samples are identified and reported. However, the complete training data is not always available, and retraining the model solely on recent data is not viable. To address this challenge, we adopt an incremental learning approach. This approach enables the model to incorporate new defense patterns based on recent inputs while retaining knowledge of earlier strategies. Nonetheless, there are three major challenges in applying incremental learning to refine the detection model: catastrophic forgetting, data imbalance, and plasticity loss. In the following, we describe our strategies to address each of these challenges.

The first challenge is \emph{catastrophic forgetting}, a phenomenon in which a model's performance on previously learned tasks deteriorates when it is trained on new data using conventional methods. 
In our setting, this arises because the Shield lacks access to the full historical dataset and must be updated solely using newly reported samples. 
In particular, as the Spear continuously generates data, training with all historical data to prevent catastrophic forgetting is computationally impractical and could delay timely updates. Thus, the Shield must balance efficiency and performance.
To address this issue, \emph{replay} is a widely adopted and intuitive strategy that emulates human cognition~\cite{human_knowledge_replay_Wilson1994, human_knowledge_replay_Tambini2013} by revisiting a subset of previously encountered exemplars. In our approach, we utilize the classifier as a feature extractor by capturing the outputs from its final fully connected layer to guide exemplar selection. The goal is to ensure that the selected exemplars remain representative of the current state of the classifier.

We begin by initializing the exemplar set as empty. Prior to each training round, newly selected samples are added to the existing exemplar set, and we compute the mean of their extracted features, denoted as $\bar{\varphi}$. Next, we sort the data points $\{d_i\}$ according to the distance between their corresponding feature representations $\{\varphi_i\}$ and the mean feature $\bar{\varphi}$. A representative subset is then formed by downsampling from this sorted list to serve as the exemplar set for the upcoming round. This procedure enables efficient identification of representative samples, thereby enhancing the model's ability to retain knowledge across sequential tasks.

The second challenge is \emph{data imbalance}, caused by a disproportionate number of normal versus abnormal samples. This problem often occurs due to the characteristics of the Spear and similar tools, which tend to produce many more normal samples, making effective model training more difficult.
To mitigate this issue, we incorporate a balance penalty term into the Binary Cross-Entropy (BCE) loss, defined as:
\[
\mathcal{L} = \mathcal{L}_\text{BCE} + \frac{\lambda}{B} \cdot \left| C_{\text{normal}} - C_{\text{abnormal}} \right|
\]
Here, $\mathcal{L}_\text{BCE}$ denotes the standard binary cross-entropy loss, and $B$ is the batch size. The terms $C_{\text{normal}}$ and $C_{\text{abnormal}}$ represent the number of correct predictions for the normal and abnormal classes, respectively. The balance factor $\lambda$ adjusts the strength of the penalty, encouraging the model to maintain balanced performance across both classes.

\input{algorithm/continuous-backpropagation}

The last challenge in incremental learning is the \emph{loss of plasticity}, which refers to the model’s declining ability to adapt to new information over time. During extended training, we observed that the model initially exhibited steady performance improvements. However, in later stages, its performance deteriorated significantly, indicating a loss of plasticity—i.e., the model's ability to adapt to new information diminished over time. This phenomenon suggests that standard deep learning methods struggle to maintain adaptability under prolonged training. To preserve the model’s plasticity, we adopt a strategy known as \emph{continual backpropagation}, a machine learning technique designed to sustain a network's adaptability during incremental learning. Theoretical studies~\cite{Loss-of-plasticity-in-deep-continual-learning} have shown that the plasticity of neural networks tends to decline throughout the course of stochastic gradient descent (SGD), as certain neurons become inactive and cease contributing meaningfully to the learning process.

The key idea of continual backpropagation is to identify and reinitialize such inactive neurons dynamically during training, thereby restoring the network's plasticity. Algorithm~\ref{algo:continuous-backpropagation} outlines the steps of this procedure. In each training iteration, we estimate each neuron's contribution to the network by evaluating its activation state and updating a corresponding utility value (Algo~\ref{algo:continuous-backpropagation}, line 6). This utility is accumulated from the neuron's hidden outputs: if a neuron consistently produces low activation values, its outputs are unlikely to influence subsequent layers, indicating low contribution to the model’s predictions.

After multiple updates, neurons with significantly lower contribution scores $\mathbf{u}_{i,j}$ are identified as \emph{inactive}. The decay rate $\eta$ is used to compute the running average of these contributions, ensuring stability in measurement over time. Meanwhile, we monitor the number of inactive neurons (Algo~\ref{algo:continuous-backpropagation}, line 8), which serves as a trigger for the neuron replacement mechanism. The replacement rate $\rho$—typically set to a small value—controls the frequency of neuron resets, such that only a single neuron is replaced after hundreds of updates. When the replacement is triggered, the neuron with the lowest utility is selected and its input and output weights are reinitialized. In particular, the output weights are reset to zero to prevent immediate downstream influence.

This approach enables the continual monitoring of network activity and the adaptive replacement of underperforming units. By injecting controlled randomness and non-gradient-based updates, we maintain the model’s flexibility and responsiveness to new data. Importantly, the low replacement rate ensures that the overall network architecture remains stable, while the neuron configuration is gradually refined over time.

\emph{Implementation for TE.}
We begin by randomly initializing a binary Multilayer Perceptron (MLP) as the detector and setting the exemplar set to empty. 
In each evolution round, the detector attempts to detect attacks. If detection fails, the misclassified data is collected and forwarded to the Shield for refinement. We then construct a training set by combining the misclassified samples with the existing exemplar set, and retrain the Shield using the loss function $\mathcal{L}$ alongside continual backpropagation, incorporating an early stopping criterion to obtain the updated detector.

Following training, the updated detector is employed as a feature extractor to select a new exemplar set from the current training data. This evolution process is repeated until the detector consistently achieves detection performance above a predefined threshold.

%% file: algorithm/algorithm-gen-attack-vector.tex
\IncMargin{1em}
\begin{algorithm}[t]
\footnotesize
\SetKwData{Left}{left}\SetKwData{This}{this}\SetKwData{Up}{up}
\SetKwFunction{Union}{Union}\SetKwFunction{FindCompress}{FindCompress}
\SetKwFunction{GeneratePopulation}{GeneratePopulation}
\SetKwInOut{Input}{input}\SetKwInOut{Output}{output}
\SetKwInOut{Parameter}{parameter}
\SetKwProg{Fn}{Function}{}{}

\SetKwComment{tcp}{// }{}

    \Input{Vector of current CPS status $v_0$, prediction model $M_p$, max feature number $n_m$, population size $p$, fitness function $f$}
    \Parameter{Boolean flag $useGA$ to enable genetic algorithm, number of generations $g$ (if $useGA$ is true)}
	\Output{Attack vector $v_a$}
    \BlankLine 
    
    Let $V_s$ := $\langle \rangle$ \tcp*[l]{Initialize sequence}
    \While{Size of $\ V_s < p$}{
        Construct an attack vector $v$ from $v_0$ by randomly selecting and manipulating $n \le n_m$ features\;
        $s_p := M_p(v)$\;
        $V_s := V_s \cup \langle v_p, f(s_p) \rangle$\;
    }
    
    \tcp{Apply genetic algorithm if required}
    \If{$useGA$}{
        \For{$gen \gets 1$ \KwTo $g$}{
            $V_s := V_s\ \cup $ CrossoverAndMutate($V_s$) \;
            $V_s := $ Select a subset from $V_s$ using \textit{Roulette Wheel Selection} based on $v$'s fitness value $f(M_p(v))$, where $v \in V_s$\;
        }
    }
    
    Select an attack vector $v_a$ from \textit{V} using \textit{Roulette Wheel Selection} with corresponding fitness values $f(s_p)$\;
    \Return{$v_a$}\;

    \caption{Attack Vector Generation with Optional Genetic Algorithm}
    \label{algo:gen-attack-vector}
\end{algorithm}
\DecMargin{1em}

%% file: algorithm/coverage-guided.tex
\IncMargin{1em}
\begin{algorithm}[t]
\footnotesize
\SetKwData{Left}{left}\SetKwData{This}{this}\SetKwData{Up}{up} \SetKwFunction{Union}{Union}\SetKwFunction{FindCompress}{FindCompress} \SetKwInOut{Input}{input}\SetKwInOut{Output}{output}
\SetKwComment{tcp}{// }{}
	
    \Input{Candidate attack embeddings $E=\langle e_1, e_2, ..., e_n \rangle$, past attack embeddings $F=\langle f_1, f_2, ..., f_m \rangle$, feature extractor $\varphi$ from predictor}
	\Output{Rating for candidate attacks $D'$}
    \BlankLine 
    $D \leftarrow \emptyset$
    
    \For{each attack $e_i$ in $E$} {
        $d_i = \frac{1}{m}\sum_{j=1\sim m}(\texttt{l2-norm}(e_i, f_j))$
        
        $D = D \cup d_i$
    }
    $d_{max}, d_{min}$ = \texttt{max}($D$), \texttt{min}($D$)
    
    normalize $d$s in $D$ with $d_{max}$ and $d_{min}$ using \textbf{Min-Max Scaling} to get $D'$
    \caption{Coverage-guided Fitness Function}
    \label{algo:coverage-guided}
    \end{algorithm}
\DecMargin{1em}

%% file: algorithm/continuous-backpropagation.tex
\IncMargin{1em}
\begin{algorithm}[t]
\footnotesize
\SetKwData{Left}{left}\SetKwData{This}{this}\SetKwData{Up}{up} \SetKwFunction{Union}{Union}\SetKwFunction{FindCompress}{FindCompress} \SetKwInOut{Input}{input}\SetKwInOut{Output}{output}
\SetKwComment{tcp}{// }{}
	
    \Input{Replacement rate $\rho$, decay rate $\eta$, $l$-layer DNN defender $D$}
	\Output{new defender $D'$}
    \BlankLine 
    Initialize utilities $\textbf{u}_{i,j} = 0$ for each neuron $n_{ij} \in D$.
    Initialize replace flag $\textbf{c}_{j} = 0$ for each layer $j$ in $D$.
    
    \For{each data point $\textbf{x}$} {
        Pass $\textit{\textbf{x}}$ through the network for forward propagation.
        
        \For{each layer $i$ in $D$} {
            \For{each $neuron_{i,j}$ in layer $j$} {
                $\textbf{u}_{i,j} = \eta \times \textbf{u}_{i,j} + (1 - \eta) \times |h_{i, j}|$ \tcp*[r]{$h_{i, j}$ is the value of neuron $j$ in layer $i$  }  
                Find the number of inactive neurons: $\textbf{\textit{n}}_{\text{inactive}}$ in layer $j$ that $h_{i, j} == 0$.

                $\textbf{c}_{j} = \textbf{c}_{j} + \textbf{\textit{n}}_{\text{inactive}} \times \rho$
            }
            \If{$\textbf{c}_{j} > 1$} {
                Find the $neuron_{r,j}$ in layer $j$ with the smallest utility $\textbf{u}_{r, j}$.

                Reinitialize $neuron_{r,j}$'s input weights to random numbers.

                Reinitialize $neuron_{r,j}$'s output weights to 0.

                $\textbf{u}_{r, j} = 0$

                $\textbf{c}_{j} = \textbf{c}_{j} - 1$
            }
        }
    }
    \caption{Continual Backpropagation for Training a $l$-layer DNN}
    \label{algo:continuous-backpropagation}
    \end{algorithm}
\DecMargin{1em}

%% file: content/evaluation.tex
\vspace{-0.2em}
\section{Evaluation}\label{sec:eval}
\vspace{-0.2em}
\label{sec:evaluation}
We evaluate the effectiveness and performance of our method for constructing an anomaly detector using the TE and RAAW testbeds (Section~\ref{sec:background}) by addressing the following research questions:

\begin{itemize}
    \item \textbf{RQ1 (Attack Discovery)}: How comprehensive is the Spear in discovering potential attack vectors?
    \item \textbf{RQ2 (Attack Detection)}: Can we build a robust Shield capable of accurately detecting attacks?
    \item \textbf{RQ3 (Parameter Sensitivity)}: Is \textit{Evo-Defender} robust to different anomaly detector widths and sliding window sizes?
    \item \textbf{RQ4 (Ablation Study)}: What is the contribution of each module in the Shield evolution system to its overall performance?
\end{itemize}

\input{tables/rq1}
\vspace{-0.4em}
\subsection*{RQ1: Attack Discovery}
\vspace{-0.2em}
The first RQ aims to assess the Spear's ability to discover real and meaningful attacks, thereby ensuring that the collected data can effectively enhance the anomaly detector and support reliable evaluation. In this experiment, we record all historical data generated during the attack discovery phase for both the TE and RAAW testbeds. Our goal is to establish a foundational understanding of the Spear's performance. Since the test data used in the subsequent RQs is generated by the Spear, this RQ serves to validate its effectiveness and ensures a fair and meaningful evaluation of the anomaly detector. 

Furthermore, to analyze the diversity of the attacks discovered by the Spear, we examine the various system states encountered by the CPS under test. We employ t-Distributed Stochastic Neighbor Embedding (t-SNE), a dimensionality reduction technique that preserves pairwise distances while projecting high-dimensional data into a lower-dimensional space, thereby facilitating a clearer interpretation of the system's behavioral patterns.

\begin{figure}[t]
    \centering
    \includegraphics[width=0.8\linewidth]{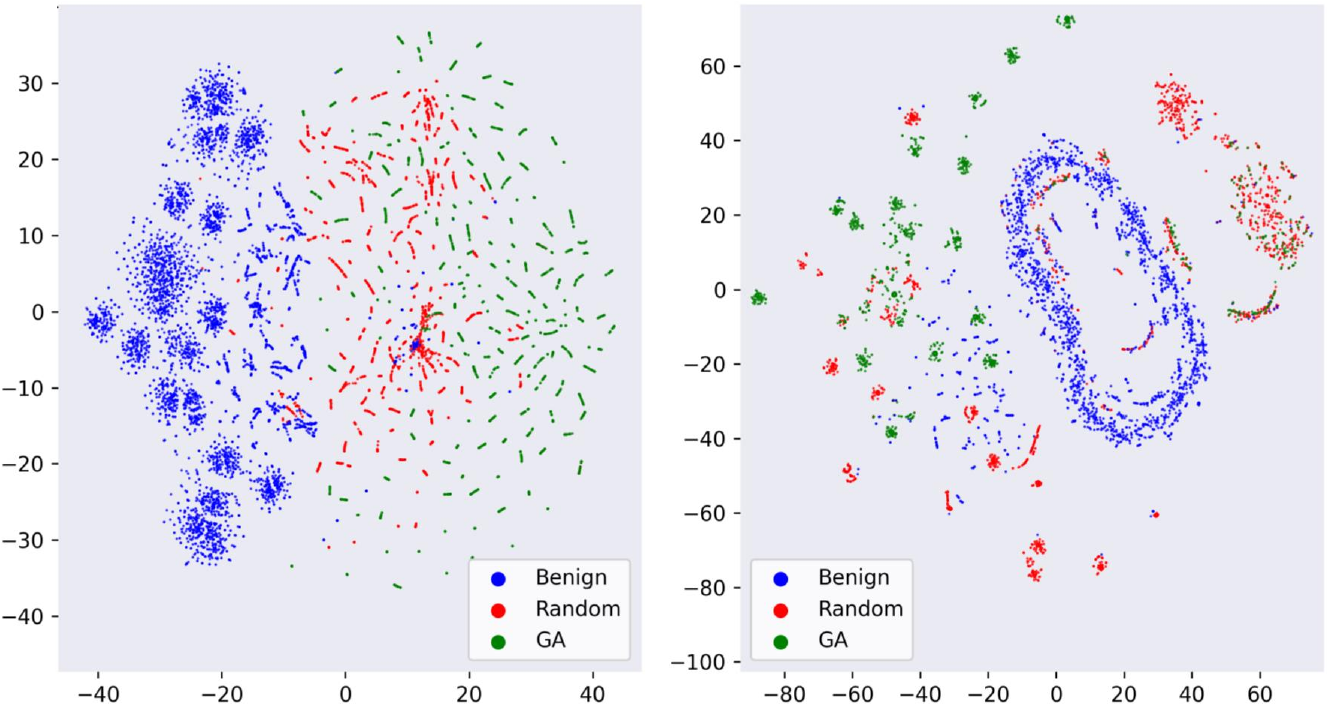}
    \vspace{-0.5em}
    \caption{Diversity Analysis of System States Using t-SNE. \textbf{Our approach pushes CPS into diverse states, uncovering complex and subtle attack vectors.}}
    \vspace{-1.5em}
    \label{fig:tsne}
\end{figure}
\textbf{\textit{Setup.} }
In our study, we conducted different attack scenario injections on the TE and RAAW platforms, dividing them into training and evaluation datasets. Each set of attack vectors was generated independently to ensure a diverse testing environment.

Specifically, for the TE platform, attack injections were initiated after 5 hours of system operation, with a maximum allowable runtime of 12 hours. Runs shorter than 5 hours were excluded from our analysis to ensure the system reached a stable state and to eliminate cases where a TE shutdown was caused by aggressive sensor drift before the attack.
Attacks that resulted in a system shutdown between the 5 and 12-hour mark were classified as effective, while those that reached the maximum runtime without triggering a shutdown were deemed ineffective.
To assess the safety constraints of the TE platform, we identified five critical sensors linked to shutdown events: reactor pressure, reactor level, reactor temperature, separator level, and stripper level.
Any readings from these sensors that exceeded predefined safety thresholds resulted in an immediate system shutdown. 

In the RAAW experiment, we injected attacks during the system's operation, resulting in three potential outcomes. It is worth noting that for the robotic arm, it is straightforward to reset to the normal condition.
The first scenario involved an ineffective attack, allowing the system to complete the stacking process normally. The second scenario featured an immediately detected effective attack, causing an instant pause, such as a spoofed access control sensor triggering an internal alarm. The third scenario involved an undetected effective attack that disrupted the system's workflow, persisting until the runtime exceeded the predefined operational limit without completing the stacking task.
Different attack strategies may lead to the same sequence of visited states. To explore the diversity of abnormal states identified by the Spear, we sampled the training data to assess the states visited by the CPS under test. We ensured that the sampling length matched the anomaly detector's time interval. Each sampling point represents the CPS state over a specific time interval, formatted as a two-dimensional tensor.

\textbf{\textit{Result.} }
Our experiments on the TE and RAAW platforms demonstrate the effectiveness of the Spear's ability to execute successful attacks in the absence of any anomaly detector (Tables \ref{tab:rq1-merged-raaw-te}). Specifically, on the TE platform, we collected 223 attacks for the training set and 113 for the test set, achieving an attack success rate exceeding 90\%. 
Notably, more than half of these successful attacks caused one or two sensors to enter abnormal states, with attacks triggering two sensors more frequently than those affecting only one. 
An analysis of the trigger patterns revealed that all sensors were activated multiple times, except for the \textit{Reactor Temperature}, which did not directly trigger any anomalies. 
Nevertheless, we identified 14 traces across the dataset in which the reactor temperature increased; in each case, the rise indirectly caused other sensors to exhibit abnormal behavior. 
This phenomenon occurs because even a slight temperature increase in the TE system can propagate and impact other subsystems, violating safety constraints before the temperature itself crosses the abnormal threshold.

In the RAAW dataset, we injected 200 attacks into the training set and 100 into the test set, with over three-quarters of the attack vectors resulting in successful attacks. 
Among these, more than half evaded detection by the built-in anomaly detector and disrupted normal operations. 
For instance, one successful attack set the variable \texttt{BLDC\_01\_Auto\_FWD\_Start} to 1, causing the conveyor motor to remain continuously active. 
This led to a misalignment of plastic blocks being pushed from the hopper onto the conveyor, ultimately resulting in material jams that halted the production process.

Figure~\ref{fig:tsne} presents the t-SNE analysis results for data collected from both platforms.  
On the TE platform, a clear separation is observed between blue (normal) points and green and red (abnormal) points. 
Samples generated using the GA successfully explored novel regions of the state space that were not covered by the random algorithm. 
Samples from the same time series tend to form compact clusters, with abnormal samples within a cluster often exhibiting linear distributions. 
This is primarily due to two factors: the relatively small number of malicious samples necessitated the inclusion of all abnormal points, and the dominance of floating-point features in the TE system resulted in smooth variations across samples.

Similarly, on the RAAW platform, normal and abnormal states are distinctly separated, and samples from the same sequence cluster together. 
However, due to the prevalence of discrete features in the RAAW dataset, transitions between states exhibit less continuity. 
Jumps in discrete variables reduce the similarity between consecutive samples, leading to more fragmented and independent clusters.

Overall, attacks generated by the random algorithm tend to trigger states close to the nominal operating conditions, while those generated by the GA algorithm explore a broader and more diverse range of scenarios. 
These results highlight the differences in how attack algorithms traverse the system's state space and underscore the superior effectiveness of the GA algorithm in uncovering complex and subtle attack vectors.

\input{tables/rq2.tex}
\vspace{-0.4em}
\subsection*{RQ2: Attack Detection}
\vspace{-0.2em}

For this RQ, our goal is to assess the ability of our defender to handle real world attacks mentioned in \textbf{RQ1}. For comparison, we select prior CPS anomaly detection methods \cite{anomaly-detect-unsupervised-learning-yuqi, ads-for-ics-1-esorics2022, ads-for-ics-2-Kravchik2019EfficientCA, ads-for-ics-3-Zizzo2019IntrusionDF} as baselines.
We evaluate our defender’s performance using two complementary methods: \emph{sample classification} and \emph{end-to-end testing}. 

\textbf{\textit{Setup.} }In the sample classification setting, we divide the test data from RQ1 into individual samples, each of which is a window of data consisting of consecutive actuator values and sensor readings, representing the smallest input unit for the defender.
Specifically, we segment 103 test logs on the TE platform into 1,144,305 samples of length 150 time points, and 44 logs on the RAAW platform into 17,559 samples of length 50 time points. 
We assess Shield’s performance in terms of accuracy, precision, recall, and F1 score on these samples, thereby evaluating its capability to detect localized temporal anomalies.

In end-to-end testing, we evaluate Shield’s attack detection capability using continuous logs corresponding to individual attack injections. Specifically, for the TE dataset, we divide each log into multiple segments, each consisting of 50 time points. A log is labeled as under attack only if there are at least 100 consecutive segments that are detected as anomalous. For the RAAW dataset, we set the segment length to 40 time points and consider a log to be under attack if at least 10 consecutive segments are flagged.
These thresholds are based on observed attack patterns across various systems. Their selection will be addressed in the following RQs.

This evaluation assesses the defender’s ability to identify previously encountered attacks (memorization) as well as novel, unseen attack instances (generalization). Baseline methods are evaluated on the same datasets for comparative analysis.

Together, these evaluation strategies provide a comprehensive assessment of the defender’s classification ability on discrete samples and its real-world effectiveness on continuous operational data, validating both its detection robustness and its stability against sustained attacks.

\textbf{\textit{Result.} }
Table~\ref{tab:rq2-sample-classification-raaw-te} presents the sample classification performance of various detection models on the TE and RAAW datasets. Across both datasets, our method consistently outperforms all baselines. On the TE dataset, Evo-MLP achieves the highest accuracy (92.03\%), precision (84.52\%), and F1 score (50.84\%), demonstrating a significant improvement over previous methods by yielding fewer false positives and achieving a better balance between false positives and missed detections. 
This highlights its overall effectiveness in accurately detecting relevant samples.
While deep learning baselines (CNN, GRU, LSTM) achieve recall rates exceeding 99\%, further analysis reveals their vulnerability to sensor drift. This results in increased false alarm rates, thereby reducing their overall accuracy and precision compared to our approach.

For the RAAW dataset, Evo-MLP again achieves the best overall performance, with an accuracy of 89.59\%, precision of 100.00\%, recall of 75.42\%, and an F1 score of 85.99\%. Although the CNN, GRU, and LSTM baselines attain perfect precision, their lower recall and accuracy compared to Evo-MLP indicate a more conservative behavior, leading them to miss certain attack instances.

It is also noteworthy that the baselines demonstrate stronger recall on the TE platform, while exhibiting much higher precision on the RAAW platform. This discrepancy can be attributed to the characteristics of the variables in each testbed. In the TE testbed, many variables are continuous, such as temperature readings, which can range from tens to hundreds of degrees. In contrast, most variables in the RAAW testbed are discrete, representing operational states (e.g., the value of a laser detector is either 0 or 1).
As a result, when calculating the difference between predicted and actual values, the prediction model in TE is more likely to produce larger values, potentially exceeding the detection threshold, leading to higher false alarm rates. On the other hand, in the RAAW testbed, abnormal states may manifest in only a few key variables, making it difficult for the loss function to capture the anomaly effectively. This, in turn, increases the likelihood of missed attacks.

Table \ref{tab:rq2-end-2-end-raaw-te} summarizes the end-to-end detection performance of various defender models on both platforms. 
For seen attacks on the TE dataset, Evo-MLP achieved a 97.1\% detection rate with a moderate 10.2\% false alarm rate. Predictor-based defenders (CNN/GRU/LSTM) detected all attacks (100\% detection success) but suffered from high false alarm rates, ranging from 34.6\% to 35.1\%. On unseen attacks in the TE dataset, Evo-MLP maintained a strong detection rate of 86.4\% with a low false alarm rate of 6.8\%, while predictor-based defenders continued to achieve perfect detection but triggered high false alarms (35\%). 
These results are consistent with the sample classification performance. They also highlight that predict-based approaches are overly sensitive when applied to chemical systems with numerous continuous sensors.

For the RAAW dataset, Evo-MLP achieved 93.7\% detection success against known attacks with only 7.4\% false alarms. Predictor-based defenders struggled, with CNN achieving only 11.6\% success and LSTM performing the best at 36.8\%. For unseen RAAW attacks, Evo-MLP continued to perform strongly, achieving 86.4\% detection with no false alarms, highlighting its superior generalization capability.

These results show that Evo-MLP maintains strong detection performance and low false alarm rates across datasets and attack scenarios—crucial for CPS operations. 
In contrast, predictor-based detectors consistently fail, either missing attacks or generating false alarms due to fundamental weaknesses in handling attack patterns and CPS characteristics.
This is also evidenced by their low recall in sample classification experiments, highlighting their high false negatives and limited effectiveness in detecting attacks.

Overall, the comprehensive evaluation shows that our approach outperforms predictor-based baselines in both sample classification and end-to-end detection, consistently achieving high detection rates with low false alarm rates across diverse CPS datasets and attack scenarios.
Its strong performance against both known and novel attacks highlights its adaptability and reliability for real-world deployment.

\begin{figure}[t]
    \centering
    \begin{subfigure}[t]{\linewidth}
        \centering
        \includegraphics[width=0.8\linewidth]{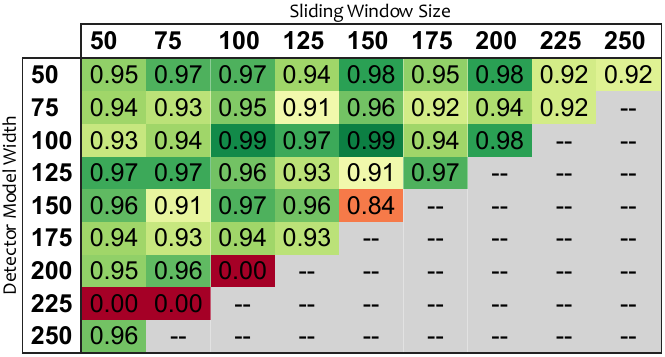}
        \caption{TE}
        \label{fig:rq3-te}
    \end{subfigure}
    
    \vspace{0.5em}

    \begin{subfigure}[t]{\linewidth}
        \centering
        \includegraphics[width=0.64\linewidth]{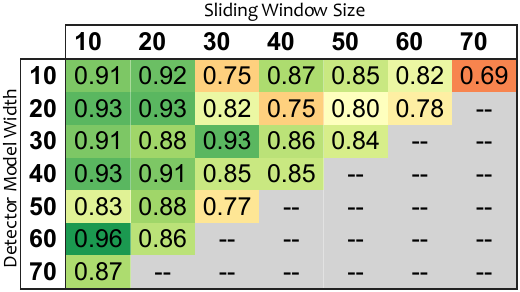}
        \caption{RAAW}
        \label{fig:rq3-raaw}
    \end{subfigure}
    
    \caption{Analysis of Anomaly Detector Width and Sliding Window Size. The vertical axis shows the detector model width, while the horizontal axis shows the sliding window size. Performance is displayed as both numeric values and color, with green indicating higher accuracy and red indicating lower accuracy. \textbf{Our method allows flexible selection of anomaly detector width and sliding window size, with most settings yielding good performance.} }
    \vspace{-1em}
    \label{fig:rq3-subgraphs}
\end{figure}

\vspace{-0.4em}
\subsection*{RQ3: Parameter Sensitivity}
\vspace{-0.2em}
For this RQ, we aim to show that the widths of the anomaly detector and the sliding window size do not significantly impact the detection performance. These parameters are set based on empirical observations of attack patterns in different systems and experimental results. For the TE dataset, the minimum number of samples from attack injection to system shutdown is 346, so the sum of the anomaly detector’s window width and the sliding window size should not exceed this value. In the RAAW dataset, about 480 samples are generated during all stacking operations, with each plastic block producing an average of 80 time units of data, thus requiring a smaller window size. 

\textbf{\textit{Setup.}}
This experiment was conducted on both platforms, with parameter choices tailored to each platform's characteristics. For the TE dataset, the width of the anomaly detector model and the sliding window size each ranged from 50 to 250, while for RAAW, both parameters were set between 10 and 70. Since there is a theoretical upper limit for the sum of these two parameters (as described earlier), parameter combinations in the lower right corner of Figures \ref{fig:rq3-te}-\ref{fig:rq3-raaw} were not tested. To evaluate the detection capabilities and memory of each defender, we performed performance assessments on the validation set using the same approach as in RQ2.

\textbf{\textit{Result.}}
The results in Figure \ref{fig:rq3-subgraphs} show that our method is robust to most settings. On the TE dataset, detection accuracy exceeded 0.9 for most configurations, with the highest accuracy observed when the anomaly detector width was 100 and the sliding window size was 100 or 150. 
Certain parameter combinations (e.g., 200-100, 225-50, 225-75) consistently failed to converge during training, as confirmed by repeated experiments and log analysis. This suggests these settings compromised data quality, hindering effective defender evolution. As these cases approach the system's theoretical limits, occasional poor performance is expected and does not affect the overall robustness of our method.
On the RAAW dataset, most defender configurations achieved performance above 0.8, with the best results when the width was 60 and the sliding window size was 10. We also found that larger window sizes in RAAW tended to result in slightly lower detection performance.

Overall, comprehensive evaluation indicates that the detection performance of our method is insensitive to the width of the anomaly detector and the sliding window size. This demonstrates that \textit{Evo-Defender} maintains stable performance across different platforms and a variety of parameter configurations, highlighting its strong generalizability.

\begin{figure*}[t]
    \centering
    \begin{subfigure}[t]{\linewidth}
        \centering
        \includegraphics[width=0.9\linewidth]{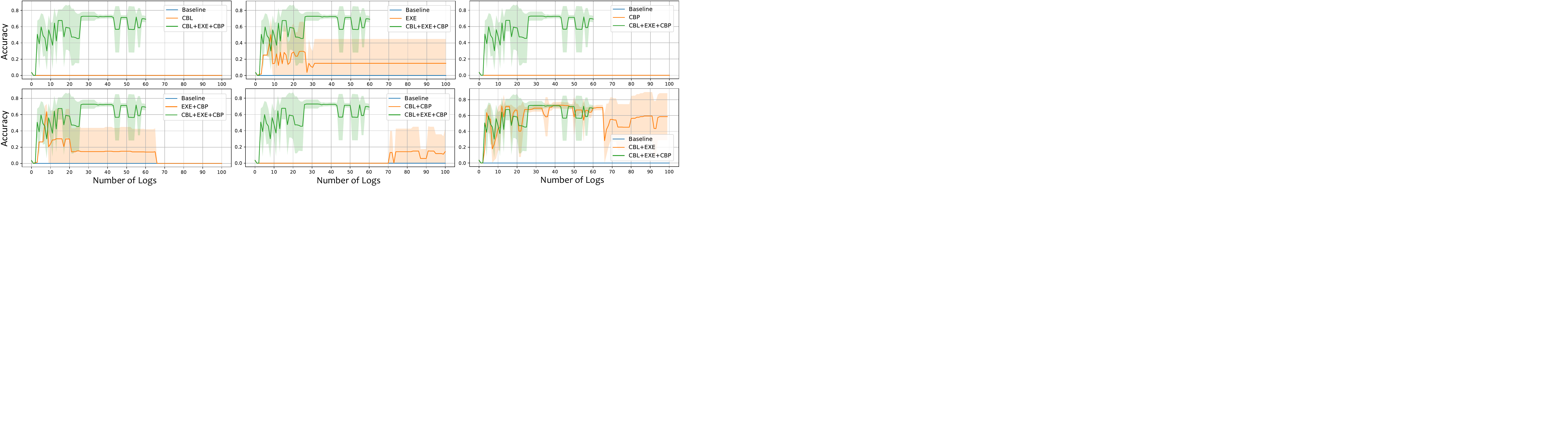}
        \caption{TE}
        \label{fig:rq4-te}
    \end{subfigure}
    
    \vspace{0em}

    \begin{subfigure}[t]{\linewidth}
        \centering
        \includegraphics[width=0.9\linewidth]{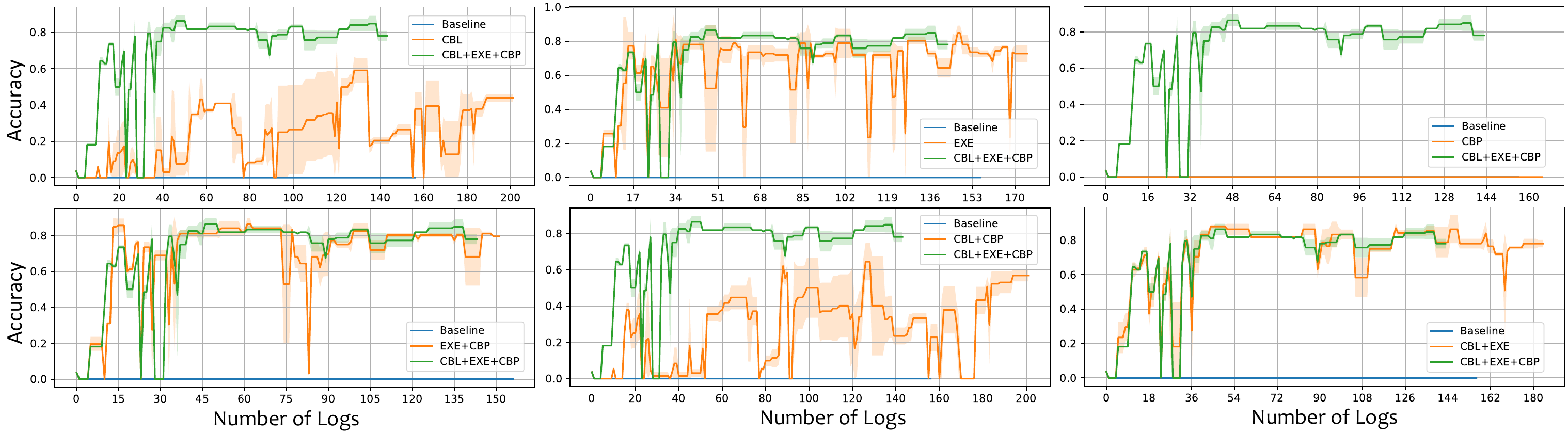}
        \caption{RAAW}
        \label{fig:rq4-raaw}
    \end{subfigure}
    
    \caption{The effect of each module on (a) TE and (b) RAAW.   
        \textbf{Overall, all module contributes positively to the evolution process, allowing the defender to reach equal or better performance with less data.} }
    \vspace{-1em}
    \label{fig:rq4-subgraphs}
\end{figure*}

\vspace{-0.4em}
\subsection*{RQ4: Ablation Study}
\vspace{-0.2em}
This research question aims to evaluate the impact of each module in the defender evolution system.
Recall that our defender evolution comprises three main modules: class balance loss, exemplar data, and continual backpropagation. 
\begin{itemize}
    \item \textbf{CBL} (Class Balance Loss): Balances the defender’s learning preference between positive and negative labels within a single round. This is particularly useful when incoming data is unevenly distributed between the two classes, thereby mitigating data imbalance issues.
    \item \textbf{EXE} (Exemplar): Using exemplars, the defender can effectively consolidate knowledge gained from historical data, reinforcing the memory of previously learned information.
    \item \textbf{CBP} (Continual Backpropagation): Incorporates continual backpropagation into the preceding modules, enhancing the defender’s adaptability and stabilizing the training process. This also helps reach the predefined stopping condition more quickly.
\end{itemize}

To systematically evaluate the contributions of the three modules in the Shield, we conducted ablation experiments on both the TE and RAAW platforms. To achieve this, we separately removed different modules of each stage and then compared their results.

\textbf{\textit{Setup.} }
Each platform evaluates eight configurations: (1) Baseline (no modules), (2-4) Single module (Baseline + Exemplar / CBL / CBP), (5-7) Pairwise combinations, and (8) Full module. 
The evolutionary process terminates when stopping conditions are met or data traces are exhausted. 
Unlike RQ2, we use a less stringent stopping condition to highlight the performance differences between modules in terms of data consumption.
We perform five independent runs for each configuration, standardize sequence lengths by truncating or padding with the final result, discard the highest and lowest values, and calculate the mean and standard deviation of the success detection rate.
Length standardization reflects the data needed for each configuration to meet the same stopping criterion, minimizing random variation between runs rather than merely truncating to the shortest length.

\textbf{\textit{Result.} }
For the TE platform, as shown in Figure \ref{fig:rq4-te}, the full-module configuration (green line) demonstrates optimal convergence, satisfying stopping criteria in 68 data logs, which is 69.6\% less than baseline (224 → 68).
While the baseline configuration suffers from catastrophic forgetting, it consistently shows lower accuracy throughout.
Among the single-module variants, EXE shows initial performance improvement, but as training progresses, it fails to learn from new data and its performance declines significantly in later stages.
Other single-module configurations perform similarly to the baseline throughout evolution.
In pairwise combinations, CBL+EXE temporarily exceeds the full module's peak accuracy but typically needs more data to meet the same stopping criteria.
The accuracy variation of the EXE+CBP combination is similar to that of the EXE single module, while CBL+CBP shows no measurable improvement.

For the RAAW platform, as shown in Figure \ref{fig:rq4-raaw}, the full-module configuration maintains its advantage with 26.28\% faster convergence than the baseline (115 vs. 156 points).
Behavioral patterns differ significantly from TE: EXE alone achieves the closest performance to the full system, though it trails by 8 percentage points and suffers more performance fluctuations, such as a sharp accuracy drop in accuracy between rounds 60 and 61.
Single-module CBL shows reduced effectiveness compared to its performance on TE, while CBP demonstrates platform-specific limitations.
The pairwise module combinations demonstrate superior effectiveness compared to individual modules on the RAAW platform.
In RAAW's composite module experiments, the combined schemes outperformed single-module approaches.
Specifically, while EXE+CBP and CBL+EXE achieved accuracy comparable to the full-module configuration, while the full module achieved the stop condition with less data and fluctuations.

Overall, all three modules provided essential support for the defender’s evolution process at varying levels. 
On TE, CBL offered the greatest performance boost, while combining EXE and CBP notably enhanced stability. 
On RAAW, the EXE module delivered the most significant performance gain, with CBL and CBP adding stability.

\subsection*{Threats to Validity}

There are some threats to the validity of our results. 
First, our method relies on predictive models, so the quality of the model may affect the effectiveness of the attack. However, we emphasize that the goal of this approach is to build robust defenders fast and stable, rather than focus on the behavior of smart attackers as in previous test suite generation work\cite{active-fuzzing}. We focus on reconstructing attack scenarios rather than simulating complex strategies.

Second, due to the lack of attack benchmarks, we use independently generated attacks to evaluate the defender. These generated attacks do not necessarily cover all possible tampering by intelligent attackers targeting different aspects of the system, so the results may not apply to certain 0-day attacks. However, our efficient and automated reconstruction of attack scenarios, as well as the robustness demonstrated under continuous adversarial attacks, strengthens our confidence in the approach.

%% file: tables/rq1.tex
\begin{table}[t]
\centering
\scriptsize
\vspace{0.1cm}
\caption{Attack Results on Training and Testing Sets for TE and RAAW}
\vspace{-1.0em}
\begin{tabular}{c|cc}
\toprule
\textbf{Dataset} & \textbf{Train} & \textbf{Test}\\
\midrule
\multicolumn{3}{c}{\textbf{TE}} \\
\midrule
\textbf{Total Number}           & 223           & 113 \\
\textbf{Attack Successful}      & 205 (91.93\%)   & 103 (91.15\%) \\
\textbf{Attack Fail}            & 17  (7.62\%)    & 10 (8.85\%) \\
\midrule
\textbf{Single}                 &   71 (31.84\%)   & 40 (35.40\%)   \\
\textbf{Double}                 &   79 (35.43\%)  &   41 (36.28\%)  \\
\textbf{Triple}                 &   47 (21.08\%)    & 19 (16.81\%)   \\
\textbf{Quadruple}                 &     8 (3.59\%)  &  3 (2.65\%)  \\
\textbf{Quintuple}          & 0 (0.00\%) & 0 (0.00\%) \\
\midrule
\textbf{Reactor Pressure}       & 110 (49.33\%)   & 49 (43.36\%) \\
\textbf{Reactor Level}          & 74  (33.18\%)   & 33 (29.20\%) \\
\textbf{Reactor Temperature}    & 0  (0.00\%)   & 0 (0.00\%) \\
\textbf{Separator Level}        & 110 (49.33\%) & 62 (54.87\%) \\
\textbf{Stripper Level}         & 108(48.43\%) & 47(41.59\%) \\
\midrule
\multicolumn{3}{c}{\textbf{RAAW}} \\
\midrule
\textbf{Total Number}           & 200             & 100 \\
\textbf{Attack Fail}            & 33  (16.50\%)   & 21 (21.00\%) \\
\textbf{Instant Alarm}          & 69 (34.50\%)    & 35 (35.00\%) \\
\textbf{Attack Successful}      & 98 (49.00\%)    & 44 (44.00\%) \\
\midrule
\bottomrule
\end{tabular}
\label{tab:rq1-merged-raaw-te}
\end{table}

%% file: tables/rq2.tex
\begin{table*}[t]
\centering
\footnotesize
\caption{Extended End-to-End Detection Results on TE and RAAW with Baseline Comparisons. \textbf{For TE, our approach achieves more robust performance, whereas the baseline has higher false positives due to over-sensitivity. For RAAW, our method greatly outperforms the baselines and shows strong generalization.}}
\vspace{-0.5em}
\begin{tabular}{c|cc|cc}
\toprule
\multirow{2}{*}{\textbf{Defender}}
    & \multicolumn{2}{c|}{\textbf{Seen Scenario}}
    & \multicolumn{2}{c}{\textbf{Unseen Scenario}} \\
\cmidrule(lr){2-3} \cmidrule(lr){4-5}
    & \textbf{Detection} & \textbf{False Alarm}
    & \textbf{Detection} & \textbf{False Alarm} \\
\midrule
\multicolumn{5}{c}{\textbf{TE}} \\
\midrule
Evo-MLP & 97.1\% (199/205) & 10.2\% (21/205) & 86.4\% (89/103) & 6.8\% (7/103) \\
CNN\cite{ads-for-ics-1-esorics2022} & 100\% (205/205) & 35.1\% (72/205) & 100.0\% (103/103) & 36.9\% (38/103) \\
GRU\cite{ads-for-ics-2-Kravchik2019EfficientCA} & 100\% (205/205) & 35.1\% (72/205) & 100.0\% (103/103) & 34.9\% (36/103) \\
LSTM\cite{anomaly-detect-unsupervised-learning-yuqi, ads-for-ics-3-Zizzo2019IntrusionDF} & 100\% (205/205) & 34.6\% (71/205) & 100.0\% (103/103) & 34.9\% (36/103) \\
\midrule
\multicolumn{5}{c}{\textbf{RAAW}} \\
\midrule
Evo-MLP & 93.7\% (89/95) & 7.4\% (7/95) & 86.4\% (38/44) & 0.0\% (0/44) \\
CNN\cite{ads-for-ics-1-esorics2022} & 11.6\% (11/95) & 0.0\% (0/95) & 18.2\% (8/44) & 0.0\% (0/44) \\
GRU\cite{ads-for-ics-2-Kravchik2019EfficientCA} & 9.5\% (9/95) & 0.0\% (0/95) & 11.4\% (5/44) & 0.0\% (0/44) \\
LSTM\cite{anomaly-detect-unsupervised-learning-yuqi, ads-for-ics-3-Zizzo2019IntrusionDF} & 36.8\% (35/95) & 0.0\% (0/95) & 31.8\% (14/44) & 0.0\% (0/44) \\
\bottomrule
\end{tabular}
\vspace{-1.5em}
\label{tab:rq2-end-2-end-raaw-te}
\end{table*}

\begin{table}[t]
\centering
\footnotesize
\vspace{0.1cm}
\caption{Sample Classification Performance on TE and RAAW. 
    \textbf{In general, our approach outperform baselines in most metrics.} }
\vspace{-0.5em}
\begin{tabular}{c|cccc}
\toprule
\textbf{Defender} & \textbf{Accuracy} & \textbf{Precision} & \textbf{Recall} & \textbf{F1} \\
\midrule
\multicolumn{5}{c}{\textbf{TE}} \\
\midrule
Evo-MLP & \textbf{92.03\%} & \textbf{84.52\%} & 36.36\% & \textbf{50.84\%} \\
CNN\cite{ads-for-ics-1-esorics2022} & 68.65\% & 26.09\% & \textbf{97.21\%} & 41.14\% \\
GRU\cite{ads-for-ics-2-Kravchik2019EfficientCA} & 68.68\% & 26.09\% & 97.02\% & 41.12\% \\
LSTM\cite{anomaly-detect-unsupervised-learning-yuqi, ads-for-ics-3-Zizzo2019IntrusionDF} & 68.71\% & 26.06\% & 96.73\% & 41.06\% \\
\midrule
\multicolumn{5}{c}{\textbf{RAAW}} \\
\midrule
Evo-MLP & \textbf{89.59\%} & \textbf{100.00\%} & \textbf{75.42\%} & \textbf{85.99\%} \\
CNN\cite{ads-for-ics-1-esorics2022} & 60.52\% & 100.00\% & 7.02\% & 13.12\% \\
GRU\cite{ads-for-ics-2-Kravchik2019EfficientCA} & 60.61\% & 100.00\% & 7.22\% & 13.47\% \\
LSTM\cite{anomaly-detect-unsupervised-learning-yuqi, ads-for-ics-3-Zizzo2019IntrusionDF} & 60.70\% & 100.00\% & 7.44\% & 13.84\% \\
\midrule
\bottomrule
\end{tabular}
\vspace{-2em}
\label{tab:rq2-sample-classification-raaw-te}
\end{table}

%% file: content/related_work.tex
\vspace{-0.2em}
\section{Related Work}\label{sec:related}
\vspace{-0.2em}
In this section, we highlight some related work addressing the broader themes of this paper: ensuring the integrity of CPSs and incremental learning.

Several recent research studies have focused on detecting and preventing CPS attacks. Popular solutions include anomaly detection, analyzing logs of running states to identify suspicious events or abnormal behaviors \cite{anomaly_detection_1, anomaly_detection_2, Truth_Will_Out, anomaly_detection_4, anomaly_detection_5, anomaly_detection_6, anomaly_detection_8, anomaly_detection_9, anomaly_detection_10, anomaly_detection_11, anomaly_detection_12, anomaly-detect-unsupervised-learning-yuqi, Detecting_Cyber_Attacks_in_ICS, TABOR, Learning_Based_Anomaly_Detection_for_Industrial_Arm_Applications, High_Performance_Unsupervised_Anomaly_Detection_for_Cyber_Physical_System_Networks}; 
fingerprinting, where sensors are monitored for spoofing by analyzing time and frequency domain features from sensor and process noise \cite{fingerprint_1, fingerprint_2, fingerprint_3, fingerprint_4, fingerprint_5}; 
invariant-based checks, which continuously monitor conditions across processes and components \cite{invariant_1, invariant_2, invariant_3, invariant_4, invariant_5, invariant_6, invariant_7, invariant_8, invariant_9, invariant_10}. 
The main difference between these works and ours is the lack of adequate and continuous feedback from CPS testing and insufficient understanding of the underlying processes and control operations (e.g., for designing physical invariants); 
as a result, physics-based anomaly detectors are typically static and cannot dynamically adapt or improve in response to evolving threats. In contrast, our approach not only enables dynamic self-improvement through incremental learning, but also significantly reduces the need for domain-specific knowledge and deployment effort. 
While causality-guided testing \cite{Finding_Causally_Different_Tests} employs formal methods to discover new attacks, Evo-Defender identifies attack vectors mainly using a coverage-based fitness function.
We believe these methods can enhance our defender evolution approach; for instance, advanced machine learning models for anomaly detection might augment our incremental defender construction.

The strengths and weaknesses of different countermeasures has been the focus of various studies.
Erba and Tippenhauer \cite{No_Need_to_Know_Physics} spoof sensor values (e.g. using precomputed patterns) and are able to evade three black box anomaly detectors published at top security conferences.
Urbinaetal. \cite{Limiting_the_impact_of_stealty_attacks} evaluated several attack detection mechanisms in a comprehensive review, concluding that many of them are not limiting the impact of stealthy attacks (i.e. from attackers who have knowledge about the system’s defences), and suggest ways of mitigating this.
Our defender evolution solution tackles these issues by integrating a smart attacker and employing incremental learning to steadily improve the defender's robustness.
Cárdenas et al. \cite{A_Framework_for_Evaluating_Intrusion_Detection_Architectures} propose a general framework for assessing attack detection mechanisms, but in contrast to the previous works, focus on the business cases between different solutions. For example, they consider the cost-benefit trade-offs and attack threats associated with different methods, e.g. centralised vs. distributed.

As a testbed dedicated for cyber-security research, many different countermeasures have been developed for TE itself.
These include anomaly detectors, typically trained on the manual crafted datsets or public datasets \cite{tennessee_eastman_process_2025, Additional_Tennessee_Eastman_Process_Simulation_Data} using deep learning techniques, e.g.\cite{GAN_AD_TE, RNN_AD_TE, Fault_detection_TE_temporal_deep_learning_models}. 
Du \cite{Fault_detection_TE_temporal_deep_learning_models} implemented fingerprinting systems based on sensor and process noise for detecting stochastic faults. 
Aoudi \cite{Truth_Will_Out} determines anomalies by compressing time-series signals into a low-dimensional subspace and calculating the distance between new inputs and the subspace centroid as an anomaly score to determine anomalies.

The application of incremental learning is a vibrant area of research \cite{class_incremental_survey_1, class_incremental_survey_2}, but targets are typically focus on the class-incremental learning and image classification tasks. 
To highlight a few examples: Yu et al. \cite{Lifelong_person_re-identification} introduced a novel Knowledge Refreshing and Consolidation (KRC) framework to address catastrophic forgetting in lifelong learning, enhancing performance on both old and new tasks through bi-directional knowledge transfer and dynamic memory interaction; 
Buzzega et al. \cite{Dark_experience_for_general_continual_learning} combines rehearsal with knowledge distillation to mitigate catastrophic forgetting in incremental image classification tasks; 
These examples focus on image classification, whereas Evo-Defender trains models to classify time series data from sensor readings. 
While some studies, like Qiao's work \cite{Class_incremental_Learning_for_Time_Series} on class-incremental learning for time series, utilize incremental learning methods on public datasets, they do not address real-world dynamic systems.

%% file: content/conclusions.tex
\vspace{-0.2em}
\section{Conclusions}\label{sec:conclusion}
\vspace{-0.2em}
In this paper, we introduced \emph{Evo-Defender}, an evolutionary framework that strengthens anomaly detection in cyber-physical systems by pairing a smart attacker with a continually adapting defender.
Through extensive evaluations on both simulated and real-world CPS testbeds, we demonstrated that \emph{Evo-Defender} achieves superior detection performance, requiring less data and fewer evolution rounds than traditional approaches.
Our ablation studies further highlight the critical roles played by continual backpropagation, exemplar replay, and class balance techniques in stabilizing and enhancing defender performance.
By simulating diverse attack behaviors and incrementally evolving the defender, \emph{Evo-Defender} offers a practical and scalable solution for securing complex, data-scarce CPS environments.